\documentclass[aps,prl,twocolumn,superscriptaddress,groupedaddress]{revtex4}
\usepackage{graphics}
\usepackage[normalem]{ulem}
\usepackage{amssymb}
\usepackage{color}
\usepackage{graphicx}
\usepackage{amsthm}
\usepackage{paralist}
\usepackage{verbatim}
\usepackage{chemarr}
\usepackage{braket}

\newtheorem{theorem}{Theorem}
\newtheorem{lemma}[theorem]{Lemma}

\newtheorem{proposition}[theorem]{Proposition}
\def\EQ#1{\begin{equation}\begin{aligned}#1\end{aligned}\end{equation}}
\def\>{\rangle}
\def\<{\langle}

\newcommand{\Tr}{\operatorname{Tr}}

\newcommand{\ketbra}[2]{|#1\rangle\langle #2|}
\newcommand{\map}[1]{\mathcal{#1}}

\begin{document}

\title{
%Quantum superposition of causal orders can superactivate the quantum capacity ---not true\\
%Quantum communication with zero-capacity channels in indefinite order---maybe\\
Quantum communication in a superposition of  causal orders}
 
\author{Sina Salek}
\affiliation{Department of Computer Science, University of Oxford, Wolfson Building, Parks Road, United Kingdom}
\author{Daniel Ebler}
\affiliation{Department of Computer Science, The University of Hong Kong, Pokfulam Road, Hong Kong}
\author{Giulio Chiribella}
\affiliation{Department of Computer Science, University of Oxford, Wolfson Building, Parks Road, United Kingdom}
\affiliation{Department of Computer Science, The University of Hong Kong, Pokfulam Road, Hong Kong}
\affiliation{Canadian Institute for Advanced Research,
CIFAR Program in Quantum Information Science, Toronto, ON M5G 1Z8}

\begin{abstract}
Quantum mechanics allows for situations where the relative order between two  processes is entangled with a quantum degree of freedom. Here we show that  such entanglement can enhance the ability  to transmit quantum information over noisy communication channels.    We consider two completely dephasing  channels, which in normal conditions are unable to transmit any quantum information. We show that, when the two channels are traversed in an indefinite order, a quantum bit sent through them  has a 25\% probability to reach the receiver without any error.   For partially dephasing channels,  a similar advantage takes place deterministically: the amount of quantum information that can  travel through two channels in a superposition of orders  can be larger than the amount of quantum information that can travel through each channel individually.   
\end{abstract}

\maketitle

%\pacs{}

% \section{Introduction \label{intro}}

{\em Introduction.} At its core, information theory is a theory of resources.  In Shannon's information theory, the resources are classical: the carriers of information are classical systems, travelling along well-defined trajectories.     In quantum Shannon theory, the carriers of information are quantum particles, whose internal state can be in a superposition of the basic classical states.   Still, all information carriers are assumed to travel along well-defined trajectories. In particular, when a message is sent through a sequence of communication channels, the order in which the channels are traversed is  assumed to be well-defined. 

In principle, quantum mechanics allows  particles to  experience  multiple  processes  in a superposition of different orders  \cite{Chiribella2013}.  In these situations, the order of the processes  becomes entangled with a quantum  degree of freedom. 
 This type of entanglement, called {\em causal non-separability} \cite{oreshkov2012quantum,oreshkov2018whereabouts},  is a new resource, which provides advantages in quantum computation \cite{Chiribella2012,Araujo2014}, nonlocal games \cite{oreshkov2012quantum}, and communication complexity \cite{guerin2016exponential}.   Recently, entanglement between the order of two processes and a quantum bit has been experimentally realized with photons, using  polarization \cite{procopio2015experimental,rubino2017experimental}  and orbital angular momentum  degrees of freedom \cite{goswami2018indefinite}.    An even more radical realization could arise in a quantum gravity scenario,  where the order of two processes could be entangled with the configuration of the masses in the universe \cite{zych2017bell}.

 Recently, we addressed the extension of quantum Shannon theory to  scenarios where the order of the communication channels can be in a superposition \cite{Ebler2017}. We found out that  the correlations  between the message and the quantum system controlling the order have dramatic consequences on the ability to transmit classical data. Specifically, a receiver with access to such entanglement can decode classical messages even if they have travelled through two completely depolarizing channels, each of which replaces the input state with white noise. This effect has been tested experimentally in a new experiement \cite{Goswami2018}.
    
For classical communication,   it turns out that the advantage of superposing orders is a generic phenomenon.  Every two  channels with constant output, when traversed in a superposition of two orders, enable the transmission of some classical information.  In contrast, advantages in quantum communication  are more elusive. For instance, the example of Ref. \cite{Ebler2017} does not apply to quantum communication:  no quantum bit can be sent reliably through two completely depolarizing channels, even if their order is  indefinite. 
   It is then  natural to ask whether the superposition of orders may offer any advantage in quantum communication, or whether instead there exist some fundamental reason why the advantages are limited to classical communication.

\begin{figure}
  \centering%
      \includegraphics[width=0.38\textwidth]{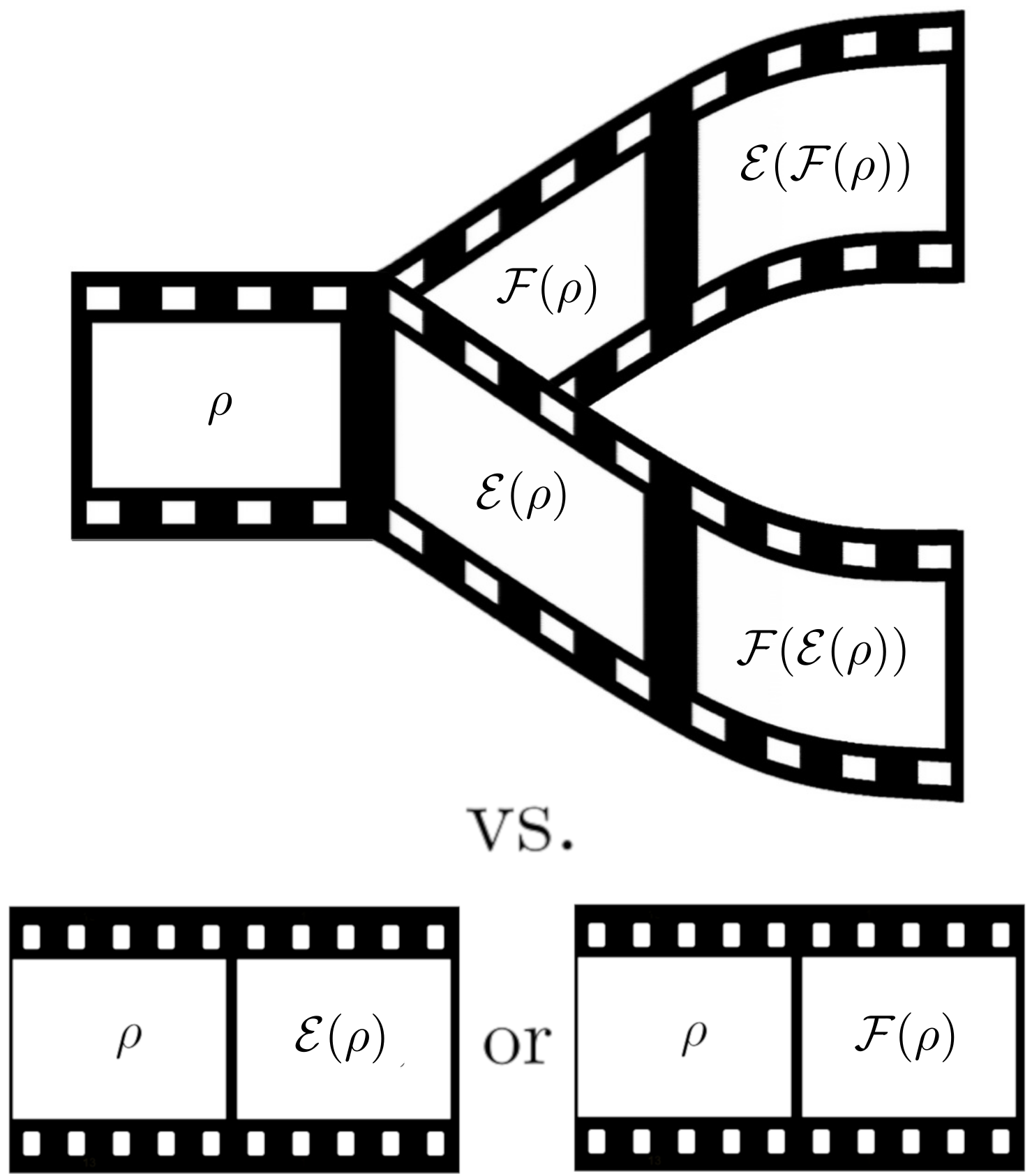}
  \caption{
  {\em On the top:} pictorial representation of the history of a quantum particle experiencing two noisy channels $\map F$ and $\map E$ in a superposition of the two alternative orders  $\map F \map E$ and $\map E \map F$. 
  Each frame shows the state of the  particle's internal degree of freedom at a given moment of time. 
  % In the last frame, the state ends up being correlated with the quantum degree of freedom that controlled the order.   
      {\em On the bottom:}  history of particles travelling through only one of the channels $\map F$ and $\map E$.  In the Letter we show that a particle travelling through the two channels in an indefinite order can carry more information than a particle travelling through each channel individually. 
    %system at each instant of time,  with the branching in the second and third frame representing the superposition of alternative histories. 
  }
    \label{diag}
\end{figure}

In this Letter we answer the question,  showing that two  noisy channels  in a superposition of two different orders can   transmit more quantum information than each channel  individually. This situation is illustrated pictorially in Figure \ref{diag}.  The simplest example involves  two  completely dephasing channels, which collapse quantum states on the two  complementary bases  $\{|0\> \, ,  |1\>\}$ and $\{|+\> , |-\>\}$, respectively. In normal conditions, no quantum information can go through either of the two channels.    Instead, as we demonstrate below, when the two channels act in a superposition of two orders, a quantum bit travelling through them has a 25\% probability to  reach the receiver {\em without any error}.  The successful transmission is heralded by the outcome of a measurement on the order qubit,  which allows the receiver to know whether the transmission has been successful or not.    
 The availability of a noiseless heralded channel can be used to implement the BB84 cryptographic protocol \cite{BennettCh1984}, or to establish entanglement in the E91 protocol \cite{ekert1991quantum}, or, more generally, to implement  any other protocol that only requires the transmission of individual qubits  uncorrelated with each other.    
 
 The combination of two channels in indefinite order also gives rise to {\em deterministic} advantages. Explicitly, we show that for partially dephasing channels there exists a parameter region where the quantum capacity of two channels in a superposition of orders is larger than the quantum capacity of each channel individually. This means that a quantum particle traversing  two channels in a superposition of two alternative orders can carry more quantum information than  % particle travelling along a definite path---in fact, the particle can carry more quantum information than 
 what each channel would  individually allow. 
 
  Our findings highlight a new link between quantum Shannon theory and indefinite causal order. In quantum Shannon theory, the ability to transmit quantum information depends on the ability to establish entanglement between the sender's and receiver's labs. Here we see that, when the  order  of the communication channels in between them is a quantum degree of freedom, it can increase the entanglement and enable transmission of quantum information beyond what was possible in a definite causal order.  

% \section{Preliminaries \label{prelim}}
% 
{\em Preliminaries.} A quantum channel $\map E$ acting on a state $\rho$ can be expressed in the Kraus  representation as $\map E(\rho)=\sum_i E_i \rho E_i^\dagger$, where $\{E_i\}$ are suitable operators.  Intuitively, each operator $E_i$ can be thought as representing an elementary process, whereby the state $\rho$ is transformed into the state $\rho_i  \propto   E_i \rho E_i^\dag$ with probability $p_i  = \Tr [  E_i^\dag E_i  \rho]$.  
 
  The quantum SWITCH \cite{Chiribella2013} describes the mechanism whereby the particle  experiences two quantum channels $\map E$ and $\map F$  in a superposition of two different orders, $\map E \map F$ and $\map F \map E$. The order depends on the state of a quantum degree of freedom, which can be taken to be two-dimensional without loss of generality.  Hereafter, we will call such degree of freedom  the {\em order qubit}.   
 Mathematically, the quantum SWITCH is a higher-order transformation \cite{chiribella2008transforming,chiribella2009theoretical}  that takes in input  two channels $\map E$ and
$\map F$, along with a state $\omega$ of the order qubit, and returns in output   the channel  $\map S_{\omega} (\map E, \map F)$ defined as
\EQ{
\map S_{\omega} (\map E, \map F)\left(\rho\right):= \sum_{i,j} W_{ij} \left(\rho \otimes \omega \right) W_{ij}^\dagger,  \label{SWITCH}
}
with Kraus operators
\EQ{
W_{ij}:=E_i F_j \otimes \ketbra{0}{0} + F_j  E_i \otimes \ketbra{1}{1} \,,  \label{SwitchKraus}
}
where $\{E_i\}$ and $\{F_j\}$ are the Kraus operators of $\map E$ and $\map F$, respectively.    Intuitively,  the Kraus operators $W_{ij}$ can be intepreted as describing a particle experiencing the elementary processes $E_i$ and $F_j$ either in the order $E_i F_j$ or in the order $F_j  E_i$, depending on whether the order qubit is in the state $|0\>$ or $|1\>$.   It is important to stress, however,  that the definition of the channel $\map S_{\omega} (\map E, \map F)$ is independent of the choice of Kraus operators for $\{E_i\}$ and $\{F_j\}$, and instead depends solely on the channels  $\map E$ and $\map F$ \cite{Chiribella2013}.

%When the order qubit is in the state $\omega=\ketbra{0}{0}$, the quantum SWITCH yields the channel $\map N_2 \circ \map N_1 \otimes |0\>\<0|$, meaning that the target system experiences channel $\map N_1$ before channel $\map N_2$ and the order is heralded by the state of the order qubit      When the order qubit is in the state  $\omega=\ketbra{1}{1}$, the quantum SWITCH yields the channel  $\map N_1 \circ \map N_2\otimes |1\>\<1|$, corresponding to the opposite order. When the order qubit is in a superposition of 
%$\ket{0}$ and $\ket{1}$, the quantum SWITCH applies the two input channels in a superposition of their
%causal order, generating a quantum correlation between the target system and the order qubit.  

In the following we will consider the use of the channel $\map S_{\omega} (\map E, \map F)$ for quantum communication between a sender and a receiver.  It is important to stress that the sender can only encode information in the particle, and not in the order qubit. The order qubit is part of the definition of the communication channel, and is only available to the receiver.  Physically, this situation would arise naturally in  the gravitational realization of the quantum SWITCH, where the order qubit corresponds to two alternative configurations of the masses in the universe \cite{zych2017bell}. In this case it is natural to assume that the sender cannot modify  the configuration of the masses, but the receiver can observe it in order to better decode the sender's message.

%It was shown that using the quantum SWITCH leads to advantages in testing properties of channels \cite{Chiribella2012,Araujo2014} and 
%communication complexity \cite{Guerin2016}, amongst others. The quantum SWITCH belongs to the broader family of so-called ``indefinite causal structures'' \cite{Oreshkov2012}.
%It has been shown, in full generality, that this resource cannot be decomposed into quantum processes with definite causal order \cite{Oreshkov2016}.

The ability of a channel $\map N$ to transmit quantum information is quantified by its quantum capacity  $Q(\map N)$, defined as the number of qubits transmitted per channel use in the limit of asymptotically many uses  \cite{Lloyd1997,Shor2002,Devetak2005}.   A lower bound to the capacity  is given by the coherent information, defined as
\EQ{
Q^{(1)}(\map N)&:=\max_\phi I(A \rangle B)_\sigma\\
&:=\max_\phi H(B)_\sigma-H(AB)_\sigma, \label{Q1}
}
where   $A$ and $B$ are the input and output systems of channel $\map N$, respectively, $H(X)_\rho=  - \Tr[ \rho \log \rho]$ is the von Neumann entropy of the system $X$ in the state $\rho$,  with $X  \in  \{  B, AB\}$,  the optimisation is with respect to all pure bipartite states $\phi_{AA'}$, involving the input system $A$ and a reference system isomorphic to $A$, and  $\sigma:=  (\map I_A \otimes \map N_{A'\to B})  \, (\phi_{AA'})$. 
%If multiple uses of a channel is allowed, the overall quantum capacity, $Q(\map N)$, given by regularisation of the coherent information, is lower bounded by the single-use capacity of the channel due to the fact that the 
Since the coherent information is generally non-additive \cite{divincenzo1998quantum}, the bound $Q (\map N)  \ge  Q^{(1)} (\map N)$ can be a strict inequality, indicating that the channel $\map N$ allows for higher capacity if  entangled inputs are used.

%\section{Main results}
{\em  Heralded noiseless communication through  dephasing channels.}   Let  $\map E$  be the bit flip channel  $\map E (\rho)   =   (1-p)\, \rho  +  p  X  \rho X$ and let $\map F$  be the phase flip channel  $\map F  ( \rho )  =   (1-q)    \, \rho +   q  Z  \rho Z$, where $X$ and $Z$ are the Pauli matrices $X =  |0\>\<1|  +  |1\>\<0|$ and $Z  =  |0\>\<0|  -  |1\>\<1|  $, and $p$ and $q$ are probabilities. 
 The bit  flip and phase flip channels belong to the 
family of  degradable channels, whose quantum capacity is exactly equal to the coherent information, namely $Q(\map N)  =  Q^{(1)}(\map N)$  \cite{Devetak2005a}. Specifically, their capacities are \cite{Wilde2013} 
\EQ{
Q(\map E)&=  1-H_2(p) \\ &{\rm and} \\    Q (\map F)    &=   1 -  H_2 (q) \, ,\label{flipCap}
}
where $H_2(x):=-x \log x-(1-x)\log (1-x)$ is the binary entropy. 

When two noisy channels are combined in a definite order, the transmission of quantum information satisfies a bottleneck inequality \cite{wilde2013quantum}, stating that the overall quantum capacity is smaller than the minimum among $Q (\map E)$ and $Q(\map F)$.    We now show that, when the two channels are combined in a superposition of orders, the bottleneck inequality can be violated. 
To this purpose, insert the bit flip and phase flip channels into the quantum SWITCH, initializing the order qubit in the state  $\omega=|+\>\<+|$. 
The resulting channel $\map S_{|+\>\<+|}(\map E, \map F)$ can be computed by separately evaluating the four terms corresponding to the operator basis  $\{|0\>\<0| \,,  |1\>\<1| \, , |0\>\<1|\, , |1\>\<0|\}$ in the space of the order qubit. 
The diagonal terms  are 
\EQ{
\label{diagonal}
\map S_{|k\>\<k|}(\map E, \map F)(\rho) =& \Big[ (1-p)  \,(1-q)\,  \rho  \\
  & + p (1-q)  X \rho  X \\
 & + q (1-p)  Z \rho  Z \\
 &+ p\, q\,   Y \rho  Y \Big ] \otimes  |k\>\<k|  \, , 
 }
where    $Y$ is the Pauli matrix $ Y = i (\ketbra{1}{0}- \ketbra{0}{1})$ and  $k  \in  \{0,1\}$. The off-diagonal terms, representing the interference between the two causal orders,    are 
\EQ{
\label{offDiagonal}
\map S_{|k\>\<k\oplus 1|}(\map E, \map F)(\rho) =& \Big[ (1-p)\, (1-q)\,  \rho  \\
 & + p (1-q)  X \rho  X \\
 & + q (1-p)  Z \rho  Z \\
 &-  p\, q\,   Y \rho  Y \Big ] \otimes  |k\>\<k\oplus 1|  \, , 
}
where $\oplus$ denotes addition modulo 2. On the particle's Hilbert space, the only difference between diagonal and off-diagonal terms is the minus sign appearing in front of the term $Y \rho  Y$.  This change of sign  arises from  the non-commutativity of the Pauli operators, expressed by  the condition  $ Z X=- X Z=i Y$.

Let us now combine the diagonal and off-diagonal terms into  the final state of the particle and the order qubit. Using Equations (\ref{diagonal}) and (\ref{offDiagonal}) one obtains the expression
\begin{widetext}
\EQ{
\map S_{|+\>\<+|}(\map E, \map F)(\rho) =    & \Big [ (1-p)\, (1-q)\,  \rho + p (1-q)  X \rho  X + q (1-p)  Z \rho  Z  \Big]  \otimes  |+\>\<+|  + \Big[  p \, q   \,   Y \rho Y \Big] \otimes |-\>\<-|    \, .\label{PauliSwitch}
}
\end{widetext}
The above state is a mixture of two states of the quantum particle, labelled by two orthogonal states  $\{  |+\> , |-\>  \}$ of the order qubit.  If the order qubit is measured in the Fourier basis,  the outcome corresponding to the state $|-\>$ will herald the presence of the state  $\rho_-  =  Y \rho Y$.    In this  case,  happening with probability $p \, q$, the receiver can decode the sender's message without any error, just by applying the correction operation  $Y$.   

The above discussion shows that two dephasing channels combined in the quantum SWITCH yield a heralded noiseless communication channel with probability $p \, q$.    This is particularly remarkable when $p=q=1/2$, because in this case the two channels $\map E$ and $\map F$ are completely dephasing, and therefore no quantum information can be sent through either of them individually. In stark contrast, a particle that traverses the two channels in a superposition of two orders can deliver a qubit to the receiver with probability 25\%.    The ability to transmit single qubits probablistically could  be used to implement  the BB84 protocol \cite{BennettCh1984}, the distribution of entanglement in the E91 protocol \cite{ekert1991quantum}, or any other cryptographic protocol that does not require the transmission of correlated qubits.   The probabilistic nature of the transmission would reduce the key rate by a factor 4 compared to the rate noiseless protocols, but this would still be infinitely better than communicating in a definite causal order, where no key could be established at all. 

Notice that the possibility of heralded noiseless communication is not limited to the bit flip and phase flip channels, but it applies more generally to every pair of qubit dephasing channels, {\em i.e.} every pair of channels that collapse the qubit's state   in {\em some} pair of bases.  Since all these channels are unitarily equivalent to a bit flip and a phase flip, heralded noiseless communication can be obtained by adjusting the bases with suitable pre-processing and post-processing operations.

 The probabilistic removal of the noise, observed when two dephasing channels are combined in the quantum SWITCH, bears some similarity with the  technique of error filtration proposed by Gisin {\em et al} in Ref.  \cite{gisin2005error}.   Error filtration concerns the propagation of a particle  a superposition of paths, each of which is subject to an independent noise.
  By performing an interference measurement in the output, it is then possible to discard  events where the noise is more severe, thereby extracting a cleaner signal.  The noise removal implemented by the quantum SWITCH is a similar phenomenon, except that the noise in the two paths is not independent:  a particle travelling through the channel  $\map S_{\omega}(\map E, \map F)$ will experience either the  elementary process $E_i F_j$ or the elementary process $F_j E_i$, which are correlated with one another.  For dephasing channels, the presence of such correlation enables a complete removal of the noise,  an effect that cannot be observed if the processes along the different paths are independent [see our appendix].

{\em Violation of the bottleneck inequality.}  We now show that indefinite causal order also  offers a deterministic advantage.  To this purpose,  we evaluate the coherent information (\ref{Q1}) of the channel   $\map S_{|+\>\<+|}(\map E, \map F)$ and we compare it with the capacities of the individual channels $\map E$ and $\map F$.  We assume $p=q$ for simplicity, so that  the two channels $\map E$ and $\map F$ have the same quantum capacity  $Q (\map E)  = Q (\map F)=  1-H_2(p)$.   

Using the same techniques as in \cite[Appendix D]{Leditzky2018}, it can be shown that the maximally entangled state is the maximising input in Equation (\ref{Q1}) for the channel $\map S_{|+\>\<+|}(\map E, \map F)$, whenever the coherent information is non-negative. With this choice, the coherent information is 
     \EQ{
Q^{(1)}(\map S_{|+\>\<+|}(\map E, \map F))=1 + H_2(p^2) - 2 H_2(p),\label{SwitchFlipCap(1)}
}
for the values of $p$ where the expression in Eq.~(\ref{SwitchFlipCap(1)}) is non-negative, and $Q^{(1)}(\map S^{\rm c}(\map E, \map F))=0$ otherwise. 

\begin{figure}
  \centering%
      \includegraphics[width=0.40\textwidth]{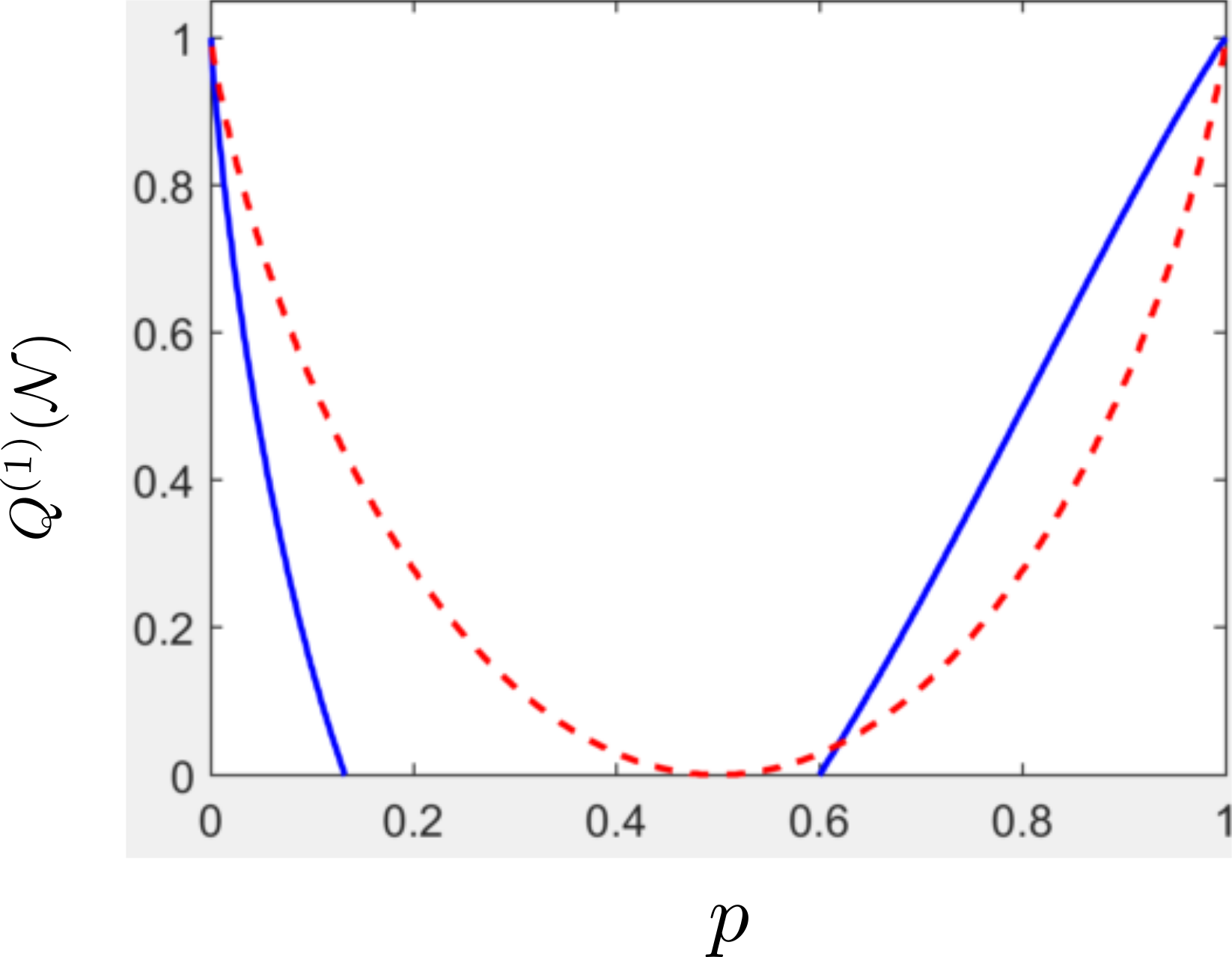}
  \caption{The blue curve shows the coherent information of bit flip and phase flip in the superposition of their causal order. The red dashed curve shows the coherent information of one bit flip channel, which is equal
  to its multiple-use quantum capacity.}
    \label{graph}
\end{figure}

%We now compare the coherent information   of the  channel  $\map S_{|+\>\<+|}(\map E, \map F)$ with the quantum capacity of the individual channels $\map E$ and $\map F$, given by  Equation (\ref{flipCap}).  
Since the coherent information is a lower bound to the capacity, every region where   $Q^{(1)}(\map S_{|+\>\<+|}(\map E, \map F))$ exceeds $Q (\map E)$ or $Q(\map F)$ will be a region where indefinite causal order offers an advantage.   The comparison is made in Figure \ref{graph}, which demonstrates the existence of an advantage  for all  values
of $p \ge 0.62$.   In this parameter region, the quantum capacity of channels $\map E$ and $\map F$ in  indefinite causal order is larger than the capacity of each channel individually. By the bottleneck inequality, this also means that the capacity in indefinite causal order is larger than the capacity of the channels $\map E \map F$ and $\map F \map E$, corresponding to definite causal orders.     In fact, the bottleneck inequality also implies an advantage over every channel of the form $\map A \map E \map B \map F \map C$, where $\map A$, $\map B$, and $\map C$ are arbitrary quantum channels matching the inputs and outputs of $\map E$ and $\map F$.

%\section{Conclusion}
{\em Conclusions}. We have shown that the possibility of indefinite causal order in quantum mechanics ha a striking impact on  quantum communication.  A  particle experiencing two processes in a quantum superposition of two alternative orders can sometimes carry more quantum information than than a particle travelling through each process individually. This holds even if both processes completely block  quantum information.  When the  processes take place in a superposition of different orders, the blockage can be lifted, allowing quantum bits to arrive intact to a receiver with a finite probability of success.   

The origin of the advantage   is that the quantum SWITCH transfers information from the initial state of the particle to the final state of the particle and the quantum degree of freedom that controlled the order.  When this happens, the quantum information is locked in the correlations: no quantum information can be retrieved from the particle alone, nor from the order-controlling system alone.    In particular,   if the order-controlling system is discarded, or projected in the basis associated to the two classical orders, then the quantum information carried by the particle becomes inaccessible.  Despite the correlations in the output state are of classical nature, they are made possible by the quantum  entanglement between the order and the order-controlling qubit, also referred to as causal non-separability \cite{oreshkov2012quantum,oreshkov2018whereabouts}.       Our work demonstrates that physical scenarios exhibiting  causal non-separability offer a resource for quantum communication, which could be exploited in key distribution protocols and other communication protocols where quantum information is encoded in single qubits. 

\medskip 

%section{Acknowledgements}
{\bf Acknowledgments.} This work is supported by the National
Natural Science Foundation of China through grant~11675136, the Croucher Foundation, the Canadian Institute for Advanced Research~(CIFAR), the Hong Research Grant Council through grant~17326616, and the Foundational
Questions Institute through grant~FQXi-RFP3-1325. This publication was made possible through the support of a grant from the John Templeton Foundation. The opinions expressed in this publication are those of the
authors and do not necessarily reflect the views of the John Templeton Foundation. 

{\small
% \bibliography{STNLN.bib}
% \bibliographystyle{plain}

}  
\appendix
\section{Comparison with error filtration}\label{appendix}  

Here we show a difference between  the quantum SWITCH mechanism and  the technique or error filtration proposed in Ref. \cite{gisin2005error}.
  Specifically, we show that when two random unitary channels are combined together, error filtration can reduce the error, but cannot eliminate it completely. 
 
Suppose that a quantum particle can travel through one of two regions, $R_0$ and $R_1$, in which it undergoes one of the two unitary gates  $U_0$ or $U_1$, respectively. The choice of the region is determined by the states $|0\>$ and $|1\>$ of a quantum degree of freedom, {\em e.g.} the path of the particle, acting as a control qubit.  The internal degree of freedom of the particle and the control qubit evolve through the gate 
%\begin{align}
$W  =  U_0  \otimes |0\>\<0|  +  U_1  \otimes |1\>\<1|$. 
%\end{align}   
If the control qubit is initially in the state $|\alpha\>   = \alpha_0  \,  |0\> +  \alpha_1  |1\>$ and is finally postselected in the state $|\beta\>   = \beta_0  \,  |0\>  +  \beta_1 \,  |1\>$, then the internal degree of freedom will experience the process described by the operator  \cite{Aharonov1990}.   
\begin{align}\label{A}
A_{U_0, U_1}  = \alpha_0 \overline \beta_0  \,  U_0   + \alpha_1 \overline \beta_1 \,  U_1\, .
\end{align}
Now, suppose that the unitary gates $U_0$ and $U_1$ are random, with probability distributions $p(U_0)$ and $q(U_1)$. If the initial state of the particle is $\rho$, then the final state will be 
\begin{align}\label{rhoprime}
\rho' \propto      \sum_{U_0, U_1}   \,   p(U_0)\,  q(U_1) \,    A_{U_0, U_1}\,  \rho  A_{U_0, U_1}^\dag \, ,
\end{align}
where the sum runs over the possible values of $U_0$ and $U_1$, assumed to be a finite set for simplicity.  

One may ask under which conditions the postselected evolution is unitary, meaning that the  postselected state  (\ref{rhoprime}) is of the form $U  \rho U^\dag$ for some unitary operator $U$.   
We have the following 
\begin{proposition}\label{prop:nonoiseless}
Suppose that the unitary $U_0 $ can take (at least) two distinct values, $V$ and $W$, and the unitary $U_1$ can take (at least) two distinct values, $S$ and $T$.  If at least three of the unitaries $\{ V, W, S, T\}$ are linearly independent, then there exist no pair of states $|\alpha\>$ and $|\beta\>$ such that the evolution (\ref{rhoprime}) is unitary. 
\end{proposition}  

{\bf Proof. }     Since equation (\ref{rhoprime}) is a convex decomposition of  $\rho'$,  the condition $\rho'  =  U\rho  U^\dag $ implies 
\begin{align}
 A_{U_0, U_1}\,  \rho  A_{U_0, U_1}^\dag    \propto  U \rho U^\dag \, ,
\end{align}  
for every $U_0$ and $U_1$.   Since $\rho$ is an arbitrary density matrix, the above condition holds if and only if 
\begin{align}\label{proptoU}  A_{U_0, U_1}\propto U\, .
\end{align} 

Without loss of generality, we assume that   $U_0$ can take the values $V$ and $W$, and that $U_1$ can take the value $S$, where  $\{V,W,S\}$  are linearly independent. Now, condition (\ref{proptoU}) implies 
\begin{align}
A_{  V,  S }  =  \lambda \,   A_{W,  S} 
\end{align} 
 for some proportionality constant $\lambda \in  \mathbb C$.  Using the definition (\ref{A}), the above condition can be rephrased as 
 \begin{align}
 \alpha_0 \overline \beta_0  V      +    ( \alpha_1 \overline \beta_1 )  (1-\lambda)   \,  S   =     \alpha_0 \overline \beta_0    \,   W  \, .
 \end{align}
 Since $V,W$, and $S$ are linearly independent, one must have  $ \alpha_0 \overline \beta_0   = 0$.  From Equation (\ref{A}), it follows that the output state is 
 \begin{align}
 \rho' \propto  \sum_{U_1}\,  q(U_1) \,     U_1  \, \rho \, U_1^\dag  \, .
 \end{align}
 Since the unitary $U_1$ can take two distinct values, the evolution $\rho \mapsto \rho'$ cannot be unitary. \qed 

\medskip  
Proposition \ref{prop:nonoiseless} can be applied to the  random-unitary channels $\map E   (\rho)  =  (1-p)\,   \rho  +  p  X  \rho X$ and $\map F  (\rho)  =  (1-q) \,  \rho   + q \,  Z\rho Z$, showing that no error filtration protocol can completely remove the noise from a particle travelling through channel $\map E$ or through channel $\map F$.

Similarly, Proposition \ref{prop:nonoiseless} can be applied to the channels $\map E  \map F$ and $\map F \map E$, both of which are random-unitary, and have the Pauli matrices  ${\sf Pauli}   =  \{  I, X, Y, Z\}$ as possible unitaries.    According to Proposition \ref{prop:nonoiseless}, a particle that experiences two independent unitary gates $U_0  \in {\sf Pauli}$ and $U_1  \in {\sf Pauli}$ will be subject to an unavoidable noise, even in postselection. 
This fact is not in contradiction with our result on the noiseless heralded communication enabled by the quantum SWITCH: in the quantum SWITCH,   the two alternative processes $U_0$ and $U_1$ are of the form $U_0  =   X^i  \, Z^j$ and $U_1  =  Z^j  \,  X^i $, for $i$ and $j$ chosen  in $\{0,1\}$ with suitable proabilities.  In this case, the unitaries $U_0$ and $U_1$ are not independent, and therefore Proposition \ref{prop:nonoiseless} does not apply.   This observation shows that the heralded noiseless communication is a consequence of the superposition of alternative {\em orders}, which correlates the  elementary processes experienced  by the quantum particle in its two alternative histories. 

\end{document}